\documentclass[showpacs,preprintnumbers,superscriptaddress,
               twocolumn,amsmath,amssymb,prd]{revtex4}
\usepackage[dvips]{graphicx}

\newcommand{\muq}{\mu_{\text{q}}}
\newcommand{\muI}{\mu_{\text{I}}}
\newcommand{\half}{{\textstyle\frac{1}{2}}}
\newcommand{\p}{\boldsymbol{p}}
\newcommand{\q}{\boldsymbol{q}}

\newcommand{\rb}{\boldsymbol{r}}

\begin{document}

\title{Larkin-Ovchinnikov-Fulde-Ferrell state in two-color quark matter}
\author{Kenji Fukushima}
\affiliation{RIKEN BNL Research Center, Brookhaven National
 Laboratory, Upton, New York 11973, USA}
\author{Kei Iida}
\affiliation{Department of Natural Science, Kochi University,
 Akebono-cho, Kochi 780-8520, Japan}

\begin{abstract}
 We explore the phase structure of two-color and two-flavor QCD in the
 space of the quark chemical potential $\muq$ and the isospin chemical
 potential $\muI$.  Using a mean-field model we calculate the chiral
 and diquark condensates, $\sigma$ and $\Delta$, self-consistently.
 In weak coupling and in the chiral limit, we confirm the interval of
 the isospin chemical potential, $0.71\Delta_0<\muI<0.75\Delta_0$, in
 which a single plane-wave Larkin-Ovchinnikov-Fulde-Ferrell (LOFF)
 phase is favored over isotropic superfluidity and normal quark
 matter.  The LOFF window becomes slightly wider at high density.  For
 stronger coupling with nonzero quark mass, which is relevant to
 currently available numerical simulations in lattice two-color QCD,
 the single plane-wave LOFF phase appears only at sufficiently high
 density.  The prediction obtained for the LOFF region could be tested
 with lattice since we can prove that the present system is free
 from the fermion sign problem.  We draw the energy landscape on which
 local minima corresponding to the isotropic superfluid phase and the
 LOFF phase and a local maximum corresponding to the gapless phase
 are manifest.  Our results clearly illustrate the path from the
 the unstable gapless phase down to the LOFF phase.
\end{abstract}
\preprint{RBRC-672}
\maketitle

%%%%%%%%%%   INTRODUCTION   %%%%%%%%%%

\section{INTRODUCTION}

  It is a longstanding problem to uncover the phase structure of dense
nuclear and quark matter in the low temperature and high baryon
density region because of its complexity.  From the academic point of
view, our curiosity urges us to imagine what an extreme state of cold
quark matter at asymptotic high density is like.  In fact, at density
far larger than the strange quark mass $M_s$ but still smaller than
the charm, bottom, and top quark masses, there has been established a
consensus that quark matter takes on color superconductivity in a
color-flavor locked (CFL)
manner~\cite{Rajagopal:2000wf,Alford:1998mk}.  Then, what comes next
as the density goes down?  This is an important question because, if
quark matter appeared in neutron star cores, its state would be
strongly affected by $M_s$.  One plausible candidate was considered to
be a gapless color superconducting phase~\cite{Shovkovy:2003uu}, which
is a QCD version of the Sarma phase~\cite{Sarma,Gubankova:2003uj}
partially stabilized by neutrality constraints.  Specifically the
gapless CFL (gCFL) phase~\cite{Alford:2003fq} was expected in quark
matter in the intermediate density region.  It turns out, however,
that the gapless phase is unlikely to exist in such quark matter
because of the chromomagnetic instability that develops at
sufficiently low
temperatures~\cite{Huang:2004bg,Casalbuoni:2004tb,Alford:2005qw,%
Fukushima:2005cm}.  At higher temperatures, a $u$-quark
superconducting (uSC) phase is predicted to occur as a remnant of the
gapless phase~\cite{Fukushima:2005cm,Ruster:2004eg}, and the existence
of the doubly critical point facing both the uSC phase and a $d$-quark
superconducting (dSC) phase~\cite{Iida:2003cc} seems
robust~\cite{Fukushima:2004zq,Abuki:2005ms}.

  Interestingly, the chromomagnetic instability in the gapless phase 
tends toward spontaneous generation of a total momentum $2\q$ carried
by each Cooper
pair~\cite{Fukushima:2005cm,Giannakis:2004pf,Fukushima:2006su}.  [In
addition to the chromomagnetic instability, an instability appears
with respect to inhomogeneous fluctuations in the gap magnitude,
leading the gapless state to a BCS-normal phase separation or a
BCS-normal mixed phase~\cite{Iida:2006df}.  See also
Ref.~\cite{Giannakis:2006gg}.]  In general, an inhomogeneous
superconducting phase characterized by pairing with nonvanishing $2\q$
is referred to as the LOFF phase named after
Larkin-Ovchinnikov~\cite{LO} and Fulde-Ferrell~\cite{FF}.  In the
present context, it is possible to describe a state resulting from the
same instability in several different ways:  a state with current
generation of collective
excitations~\cite{Huang:2005pv,Schafer:2005ym}, a state with gluon
condensation~\cite{Gorbar:2005rx}, and a colored LOFF
state~\cite{Fukushima:2005cm,Fukushima:2006su}.  They are all
equivalent algebraically.  In terms of the colored LOFF state, the
chromomagnetic instability is to be regarded as an instability with
respect to the spatially oscillating phase factor,
$e^{i\lambda^\alpha \q\cdot\rb}$ with the Gell-Mann matrices
$\lambda^\alpha$, of the pairing gap matrix in color space.
In the present paper which focuses on a two-color theory, we shall
generalize the notion of chromomagnetic instability to a
\textit{phase instability} that makes sense not only in a charged
superconductor but also in a neutral superfluid.

  In the LOFF phase the translational and rotational symmetries are
spontaneously broken by $\q$.  The single plane-wave LOFF phase is
only the simplest etude and in general a complicated crystal structure
should emerge.  In the context of QCD
physics~\cite{Alford:2000ze,Bowers:2001ip,Kundu:2001tt,Casalbuoni:2003wh},
the plane-wave LOFF phase, crystal structure, stability, and its
physical property have been examined mainly by means of the
high-density effective theory (HDET) and the Ginzburg-Landau (GL)
expansion in terms of the pairing
gap~\cite{Bowers:2002xr,Casalbuoni:2005zp,Rajagopal:2006ig,%
Mannarelli:2007bs}.  For the moment it is an urgent problem to clarify
the energetically favorable structure of the three-flavor LOFF phase.
Whereas the phase instability guarantees existence of the LOFF phase
with a lower energy than the gCFL phase even within the single
plane-wave ansatz, the HDET and GL approximations do not allow us to
identify the most favorable LOFF phase in full details.

  We shall here revisit the LOFF phase in two-color and two-flavor QCD
in the presence of both the quark chemical potential $\muq$ and the
isospin chemical potential $\muI$.  It was pointed out in
Ref.~\cite{Splittorff:2000mm} that the interval of $\muI$ in which the
LOFF phase occurs exists in such a system at least in weak coupling,
while it is nontrivial whether the LOFF window should survive or not
at stronger coupling with a finite quark mass introduced.  We will
make use of a mean-field model to address this issue by following a
line of argument of Ref.~\cite{Ratti:2004ra}, which looks successful
in reproducing the numerical results of lattice two-color
QCD~\cite{Kogut:2001if,Kogut:2001na,Kogut:2002cm,Muroya:2002ry,%
Muroya:2003qs,Hands:2006ve}.  

  Although it is beyond our current scope, we expect that the lattice
two-color QCD could observe a signature of the LOFF phase
numerically.  We adopt two sets of the model parameters, one of which
is relevant to currently available lattice simulations in which the
diquark condensate has been measured at $\muq\neq0$ but $\muI=0$.
This work is a first step toward identifying the LOFF state on
lattice.

  We can also mention that it is instructive to shed light on
two-color QCD as a mimic of real QCD in which the LOFF state is a
natural consequence of the phase instability in the gapless state.  In
the context of real QCD, we know the direction in which the unstable
gapless state goes and consider the LOFF state as a natural
replacement of the unstable state, but it remains to be surveyed how
one gives way to the other.  At this point, it is advantageous to
switch to two-color QCD.  In fact, simplicity inherent in two-color
QCD enables us to picture the energy landscape including various
states.  By doing so, we can get
a feeling that we are heading for the right way dictated by real QCD.

  This paper is organized as follows.  In Sec.~\ref{sec:setup} we 
mention unique features of two-color QCD, our mean-field model,
approximations to be made, and the twofold parameter choice at weak and
intermediate coupling.  Section~\ref{sec:result} is composed of four
subsections presenting numerical results from the mean-field model for
the isotropic superfluid phase, the unstable gapless phase, the LOFF 
phase, and the energy landscape, respectively.   Our main results are
summarized in Fig.~\ref{fig:loff_phase_w} for weak coupling with
massless quarks and in Fig.~\ref{fig:loff_phase_s} for intermediate
coupling accessible in the lattice setting.  We also plot the free
energy in the space of the diquark condensate and the pair momentum in
Figs.~\ref{fig:energy} and \ref{fig:contour} for the intermediate
coupling case.  Section~\ref{sec:conclusion} is devoted to our
conclusions and future perspectives.

%%%%%%%%%%   SETUP   %%%%%%%%%%

\section{SETUP}
\label{sec:setup}

  In this section, we will briefly summarize a chiral symmetry
breaking pattern inherent in two-color QCD, discuss the absence
of the fermion sign problem, and then describe a model for two-color
matter with nonzero $\muq$ and $\muI$ along with some approximations
in calculating the thermodynamic potential.  We will emphasize the
benefit from two-color nature that the model prediction is more robust
than in three-color case.

%%%   Symmetry Breaking Pattern   %%%

\subsection{Symmetry Breaking Pattern}

  Two-color QCD has a peculiar feature of chiral symmetry.  When quarks
are massless and the chemical potential is zero, chiral symmetry in
two-color and two-flavor QCD is augmented from the standard one
$\mathrm{SU}_\mathrm{L}(2) \times \mathrm{SU}_\mathrm{R}(2)$ to
$\mathrm{SU}(4)$.  This is because a doublet and an anti-doublet of the
SU(2) group are indistinguishable, which makes left-handed quarks and
right-handed antiquarks belong to the same group multiplet and hence
doubles the basis.  In fact this extra symmetry, which is often
referred to as the Pauli-G\"{u}rsey symmetry~\cite{Smilga:1994tb},
amounts to rotational symmetry among the chiral and diquark
condensates.  The spontaneous chiral symmetry breaking pattern is thus
$\mathrm{SU}(4)\to\mathrm{Sp}(4)$ in the presence of diquark
condensation.  Once a nonzero quark chemical potential sets in, i.e.,
$\muq\neq0$, the extended symmetry is reduced to standard global
symmetry as $\mathrm{SU}(4)\to\mathrm{SU}_\mathrm{L}(2)\times
\mathrm{SU}_\mathrm{R}(2)\times\mathrm{U_B}(1)$.  If quarks are
massive, the remaining symmetry is as usual
$\mathrm{SU}_\mathrm{V}(2)\times\mathrm{U_B}(1)$, which
spontaneously breaks down to $\mathrm{Sp}(2)$ once the quark chemical
potential exceeds the mass of the lowest-lying excitations carrying
nontrivial baryon number.  Because the $\mathrm{U_B}(1)$ symmetry is
spontaneously broken by the diquark condensation, the system is a
superfluid accompanied by massless Nambu-Goldstone bosons.  The
gauge symmetry is intact since the diquark condensate is gauge
invariant in this case.

  It is obvious that no chromomagnetic instability can happen in any
kind of superfluid state of this system because no finite Meissner
screening mass arises without color symmetry breaking.  In other
words, if we interpret it as the phase instability as we mentioned
above, the color singlet diquark condensate cannot have any spatially
oscillating phase in color space and thus cannot lead to any
instability involving color degrees of freedom.  In terms of the phase
instability, on the other hand, we can understand that fluctuations in
the $\mathrm{U_B}(1)$ phase in the gapless superfluid state could
result in instability.  It should be mentioned that the modes in the
$\mathrm{U_B}(1)$ phase itself in a superfluid correspond to the
massless Nambu-Goldstone bosons, and the instability would not emerge
in the phase factor $e^{i2\q\cdot\rb}$ but in the wave-vector $\q$ of
the phase, i.e., the current of the Nambu-Goldstone boson in a physics
language (see Refs.~\cite{Huang:2005pv,Schafer:2005ym} for details).

%%%   Fermion Sign Problem   %%%

\subsection{Fermion Sign Problem}
\label{sec:sign}

  In this subsection, we briefly summarize the sign problem of the
fermion determinant~\cite{Fukushima:2006uv} and make sure that
two-color QCD escapes from it for \textit{any} number of quark
flavors.  The Dirac operator in question reads
\begin{equation}
 \mathcal{M}(\muq)=\gamma_\mu D^\mu+\gamma_4\muq+m
\end{equation}
per one quark flavor in Euclidean space, where $D^\mu$ denotes the
covariant derivative containing gauge fields.  In the presence of the
isospin chemical potential $\muI$ with $u$ and $d$ flavors, the Dirac
operator is then a direct sum of two flavor sectors:
$\mathcal{M}(\mu_u)\oplus\mathcal{M}(\mu_d)$.  For the moment we shall
focus on the one-flavor $\mathcal{M}(\muq)$ to discuss the sign
problem, since $\det\mathcal{M}(\muq)\ge0$ is sufficient to claim
$\det[\mathcal{M}(\mu_u)\oplus\mathcal{M}(\mu_d)]
=\det\mathcal{M}(\mu_u)\cdot\det\mathcal{M}(\mu_d)\ge0$.

  Because the Euclidean gamma matrices $\gamma_\mu$'s are hermitian by
convention, the eigenvalue $\lambda_n$ of anti-hermitian
$\gamma_\mu D^\mu$, i.e.,
\begin{equation}
 \gamma_\mu D^\mu \psi_n=\lambda_n\psi_n \,,
\end{equation}
is pure imaginary.  The eigenstate $\gamma_5\psi_n$ has an eigenvalue
$-\lambda_n$, which in turn equals to $\lambda_n^\ast$.  The Dirac
determinant is, therefore, a product of all the paired eigenvalues
$\lambda_n+m$ and $\lambda_n^\ast+m$, which is non-negative because
$|\lambda_n+m|^2\ge0$.

  When $\muq$ is nonzero, $\gamma_4\muq$ is hermitian and thus the
eigenvalue $\lambda_n$ defined by
\begin{equation}
(\gamma_\mu D^\mu+\gamma_4\muq)\psi_n=\lambda_n\psi_n
\end{equation}
is no longer pure imaginary but complex.  Here again, $\gamma_5\psi_n$
has an eigenvalue $-\lambda_n$, but it is different from
$\lambda_n^\ast$ in this case.  Therefore, the Dirac determinant,
given by $\prod_n(\lambda_n+m)(-\lambda_n+m)$, is not necessarily
non-negative, and when it is negative for some gauge configurations,
the sign problem occurs.

  Two-color QCD is unique in the sense that one can find another
eigenstate $\sigma_2 C^{-1}\gamma_5\psi_n^\ast$ with an eigenvalue
$\lambda_n^\ast$, where $\sigma_2$ is the second Pauli matrix in color
space and $C$ represents the charge conjugation.  Obviously, the
eigenstate multiplied by $\gamma_5$ has an eigenvalue
$-\lambda_n^\ast$.  Consequently, the eigenvalues of the Dirac
operator always constitute a quartet: $\lambda_n+m$,
$-\lambda_n+m$, $\lambda_n^\ast+m$, and $-\lambda_n^\ast+m$, leading
to non-negative Dirac determinant through the relation
$|\lambda_n+m|^2|-\lambda_n+m|^2\ge0$.   We note that our argument
holds even with odd number of quark flavors, although in this case the
sign problem was supposed to occur~\cite{Splittorff:2000mm}.

  We can thus conclude that \textit{two-color QCD has no sign problem 
at finite density regardless of the number of quark flavors.}  This
implies that the Monte-Carlo simulation for two-color QCD with two
flavors is feasible in the presence of both $\muq$ and $\muI$, which
would help us elucidate the phase structure.

%%%   Model for Two-Color Quark Matter   %%%

\subsection{Model for Two-Color Quark Matter}

  We turn to a model for two-color and two-flavor quark matter within
the BCS ansatz that two particles with opposite momenta $\p$ and $-\p$
pair with a short-range interaction.  We may then assume that only two
mean-fields, namely, the chiral and diquark condensates,
\begin{equation}
 \sigma = G \langle \bar{\psi}_i^a \psi_i^a \rangle \;,\quad
 \Delta = G \,\epsilon^{ij}\,\epsilon^{ab}
  \langle \psi_i^{T a} iC\gamma_5 \psi_j^b \rangle \;,
\end{equation}
are predominant in the region $\muI<\muq$.  We denote the color and
flavor indices by $a$, $b$ and $i$, $j$, respectively.  Here, $T$
represents the transposition in the Dirac index, and $C$ is the charge
conjugate matrix to make the diquark condensate Lorentz invariant.
The dimensional coefficient $G$, which controls the strength of the
four-fermi coupling constant, is common to the chiral and diquark
condensates by virtue of the Pauli-G\"{u}rsey symmetry.  In a rough
but intuitively more understandable notation,
$\sigma\sim\langle\bar{u}u\rangle+\langle\bar{d}d\rangle$ and
$\Delta\sim\langle ud\rangle$.  Note that when $\muI>\muq$, the
predominant diquark condensate $\langle ud\rangle$ gives way to
$\langle u\bar{d}\rangle$.  In this work, however, we limit our
discussion solely to the region $\muI<\muq$.

  In the LOFF phase where a $u$-quark with the momentum $\q+\p$ and a
$d$-quark with the momentum $\q-\p$ form a Cooper pair with the total
momentum $2\q$, the diquark condensate has a spatially oscillating
phase as
\begin{equation}
 \Delta\;\to\; \Delta\, e^{i2\q\cdot\boldsymbol{r}} \;.
 \label{eq:loffansatz}
\end{equation}
It should be noted that we neglect any spin-one condensate such as
$\langle uu\rangle$, $\langle dd\rangle$, and
$\langle\bar{d}\bar{d}\rangle$ entirely in this work.  It generally
coexists with the spin-zero condensate $\langle ud\rangle$ in the LOFF
state, but is known to be smaller by one order of magnitude than the
spin-zero one~\cite{Alford:2000ze}.  We remark that the spin-one
condensates could be relevant for even larger Fermi surface separation
between $u$ and $d$ quarks~\cite{Splittorff:2000mm}.

  Then, the thermodynamic potential with the mean-field condensates
$\sigma$ and $\Delta$ and the pair momentum $\q$ can be expressed as
\begin{align}
 & \Omega(\sigma,\Delta,q;m_0,T,\muq,\muI) \notag\\
 & = -\frac{1}{2}\int^\Lambda \!\frac{d^3\p}{(2\pi)^3}
  \sum_{i=1}^{32}\biggl\{\frac{\epsilon_i(\p,\q)}{2}\! +T\ln\Bigl[
  1+e^{-\epsilon_i(\p,\q)/T}\Bigr]\biggr\} \notag\\
 & \qquad +\frac{\sigma^2+\Delta^2}{2G} \;.
\label{eq:potential}
\end{align}
Here $\Lambda$ is the cut-off parameter and $m_0$ is the current quark
mass.  After the $\p$-integration, the thermodynamic potential no
longer depends on the direction of the three-vector $\q$ but becomes a
function of its magnitude $q=|\q|$ alone.

  Expression~(\ref{eq:potential}) contains the sum of the quasi-quark
energies $\epsilon_i(\p)$, which correspond to the 32 eigenvalues of
the $32\times32$ quark Hamiltonian matrix with two colors, two
flavors, two spins, quark-antiquark, and Nambu-Gor'kov doubling.  The
remaining part is the energy shift associated with the mean-field
approximation.  The condensates, $\sigma$ and $\Delta$, share the same
coupling $G$ as implied by the Pauli-G\"{u}rsey symmetry; in the
chiral limit $m_0=0$ at $\muq=\muI=0$, the thermodynamic potential
$\Omega$ should be reduced to be a function of $\sigma^2+\Delta^2$.
This feature exemplifies a great advantage of two-color QCD over
various model studies of real QCD in which quantities affected by
chiral dynamics strongly depend on the model parameters.

  When $\Delta=\muI=0$, all quarks have the same constituent mass,
$M=m_0-\sigma$, leading to the quark and antiquark energies,
$\xi(\p)\mp \muq$ with $\xi(\p)=\sqrt{\p^2+M^2}$.  The effect of
nonzero $\muI$ is to shift $\muq$ by $+\muI$ and $-\muI$ for $u$ and
$d$ quarks, that is,
\begin{equation}
 \mu_u = \muq+\muI\,,\quad
 \mu_d = \muq-\muI\,,
\end{equation}
which results in crossing of the energy levels.  In the presence of
the diquark condensate $\Delta$, the crossing energy levels mix
together and a level repulsion occurs.  We will simply approximate the
energy levels by mixing between the $u$-quark energy
$\xi(\q+\p)-\mu_u$ and the $d$-quark energy $\xi(\q-\p)-\mu_d$.  Of
course we can also swap $\q+\p$ and $\q-\p$ but the integration over
$\p$ washes out the difference eventually.  The eigenvalues of the
following $2\times2$ matrix,
\begin{equation}
 \left[ \begin{array}{cc}
  \xi(\q+\p) - \mu_u & \Delta \\
  \Delta & -\xi(\q-\p) + \mu_d
 \end{array} \right] \;,
\end{equation}
give the quasi-quark energy dispersion relations,
\begin{equation}
 \epsilon_{\text{p}}^\pm(\p,\q) = \Biggl|\half\delta\xi(\p,\q)
  \pm\muI +\sqrt{\bigl[\bar{\xi}(\p,\q) \!- \!\muq\bigr]^2
  + \Delta^2}\Biggr|
\label{eq:dispersion}
\end{equation}
with
\begin{align}
 \delta\xi(\p,\q) &= \xi(\q+\p)-\xi(\q-\p)\;,\\
 \bar{\xi}(\p,\q) &= \half\bigl[\xi(\q+\p)+\xi(\q-\p)\bigr]\;.
\end{align}
Note that there are eightfold degeneracies for
$\epsilon_{\text{p}}^+(\p)$ and $\epsilon_{\text{p}}^-(\p)$,
respectively, which implies sixteen energy levels in total.  We should
remark that these energy dispersion relations are slightly different
from those used in Ref.~\cite{Alford:2000ze} because the gap energy
$\Delta$ is constant in our approximation instead of depending on the
relative angle of $\q+\p$ and $\q-\p$.  It is straightforward to
introduce any $\q$ dependence in $\Delta$.  We will, however, stick to
a constant $\Delta$, partly because we will confirm later that a
constant $\Delta$ is enough to reproduce results quantitatively
consistent with Ref.~\cite{Alford:2000ze} and partly because no
reliable ansatz is in hand a priori except when $M$ is zero.

  As for the antiquark contribution, dropping $\Delta$ is a good
approximation as long as $\Delta\ll M+\muq$.  Such an approximation,
however, would apparently ruin the Pauli-G\"{u}rsey symmetry which
should be present in the limit of $m_0=\muq=\muI=\q=0$.  Hence, we
shall keep $\Delta$ as well as $\sigma$ also for the antiquark energy
levels by adopting the same form of the energy dispersion relations as
the quark contribution with $-\muq$ replaced by $+\muq$:
\begin{equation}
 \epsilon_{\text{a}}^\pm(\p,\q) = \Biggl|\half\delta\xi(\p,\q)
  \pm\muI +\sqrt{\bigl[\bar{\xi}(\p,\q) \!+ \!\muq\bigr]^2
  + \Delta^2}\Biggr|
\end{equation}
with eightfold degeneracies again, although difference of
$\epsilon_{\text{a}}^\pm(\p,\q)$ from $|\xi(\q+\p)+\mu_u|$ and 
$|\xi(\q-\p)+\mu_d|$ is merely negligible.  We do not consider the
mixing between quarks and antiquarks which would take place with
nonvanishing $\langle u\bar{d}\rangle$ when $\muI>\muq$.

  Using the thermodynamic potential as specified above, we will
solve the following equations:
\begin{equation}
 \frac{\partial\Omega}{\partial\sigma} = 0 \;,\quad
 \frac{\partial\Omega}{\partial\Delta} = 0 \;,\quad
 \frac{\partial\Omega}{\partial q} = 0 \;,
\label{eq:gapeq}
\end{equation}
to obtain the self-consistent values of $\Delta$, $\sigma$, and $q$.
Now that we have come by all the necessary formulae, we will 
proceed to numerics next.

%%%%%%%%%%   RESULTS   %%%%%%%%%%

\section{results}
\label{sec:result}

  In this section we will present the numerical solutions to
Eq.~(\ref{eq:gapeq}) obtained for the isotropic superfluid phase, the
gapless phase, and the LOFF phase.  The gapless phase is unstable, but
still we will investigate the gapless solution to Eq.~(\ref{eq:gapeq})
for the later purpose of surveying the energy landscape.

  For numerical evaluations we first have to fix the model parameters,
namely, the current quark mass $m_0$, the cut-off $\Lambda$, and the
four-fermi coupling constant $G$.  We do not need to specify $\Lambda$
because we will present all the dimensional quantities in unit of
$\Lambda$.  Instead of $G$ we will use the diquark condensate at
$\muI=0$ denoted by $\Delta_0(\muq)$ to specify the interaction
strength; a larger $\Delta_0$ means a stronger coupling.  In this
paper we will take two parameter choices:
\begin{equation}
 \text{Parameter I:  }
 m_0 = 0\,,\:  \Delta_0(\muq\!=\!0.5\Lambda) = 0.05\Lambda \,,
\label{eq:weak}
\end{equation}
and
\begin{equation}
 \text{Parameter II:  }
 m_0 = 0.025\Lambda\,,\:
 \Delta_0(\muq\!=\!0.5\Lambda) = 0.5\Lambda \,.
\label{eq:inter}
\end{equation}
Hereafter we will refer to Parameter~I as the \textit{weak coupling}
parameter set since it will allow us to confirm that our approach
recovers the known properties of the LOFF phase at weak coupling.  For
Parameter~II, which will be denoted by the
\textit{intermediate coupling} parameter set, the choice of $m_0$ is
motivated by available lattice simulations of two-color QCD.  As we
can see in Fig.~\ref{fig:cond}, for Parameter II, the chiral and
diquark condensates amount to a comparable magnitude.  In the absence
of the sign problem of the Dirac determinant even with $\muI$
introduced as discussed in Sec.~\ref{sec:sign}, our results at
intermediate coupling could be readily compared with what we would
observe from lattice simulations.

%%%   Superfluid Phase   %%%

\subsection{Superfluid Phase}

\begin{figure}
 \includegraphics[width=7.5cm]{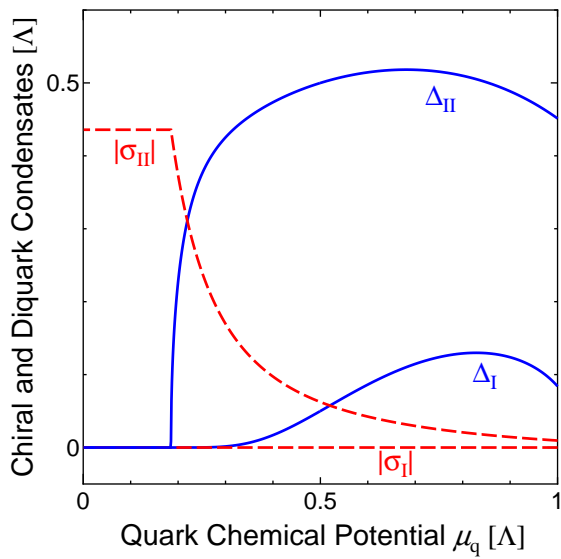}
 \caption{Chiral and diquark condensates as a function of $\muq$ at
 $\muI=\q=0$, given in unit of $\Lambda$.  The quantities with the
 subscript ``I'' and ``II'' are the results obtained at weak and 
intermediate coupling, respectively.}
 \label{fig:cond}
\end{figure}

  We show in Fig.~\ref{fig:cond} the behavior of the chiral and
diquark condensates as a function of the quark chemical potential
$\muq$ at $\muI=\q=0$.  The condensates at weak coupling with
Parameter~I~(\ref{eq:weak}) are exponentially suppressed for small
$\muq$.  This suppression is a sharp contrast to the results in the
strong coupling limit~\cite{Dagotto:1986gw,Nishida:2003uj}.  In
particular $\sigma$ keeps vanishing entirely because of $m_0=0$ as
first noted in Ref.~\cite{Dagotto:1986gw}.  From Fig.~\ref{fig:cond}
we can make sure that
$\Delta_{\text{I}}(\muq\!=\!0.5\Lambda)=0.05\Lambda$ is satisfied.  In
the case of intermediate coupling with Parameter~II~(\ref{eq:inter}),
on the other hand, the diquark condensate appears above the threshold
$\muq=0.185\Lambda$.  We can confirm that
$\Delta_{\text{II}}(\muq\!=\!0.5\Lambda)=0.5\Lambda$ as it should.
The non-monotonic behavior of $\Delta$ at $\muq\gtrsim 0.7$ is caused
by the \textit{saturation} effect:  The density cannot grow
unboundedly because the phase space is limited by the presence of
$\Lambda$, which is observed in lattice simulations as well as in model
studies~\cite{Ratti:2004ra,Nishida:2003uj}.  We remark that our
$\Delta_{\text{II}}$ and $|\sigma_{\text{II}}|$ are to be compared
with the results given in Figs.~1 and 2 of Ref.~\cite{Ratti:2004ra}.

%%%   Gapless Phase   %%%

\subsection{Gapless Phase}

  The isospin chemical potential $\muI$ plays a role in exerting
stress onto quarks to tear the Cooper pair apart.  As long as
$\muI<\Delta$, the isotropic superfluid phase is rigid against such
stress.  This can be easily understood from Eq.~(\ref{eq:dispersion})
with $\q=0$ substituted;  if $\muI<\Delta$ and $\q=0$, then
$\epsilon_{\text{p}}^++\epsilon_{\text{p}}^-=2\sqrt{(\xi-\muq)^2+\Delta^2}$,
which has no dependence on $\muI$, and thus $\Delta$ and $\sigma$ stay
constant regardless of $\muI$.

\begin{figure}
 \includegraphics[width=7.5cm]{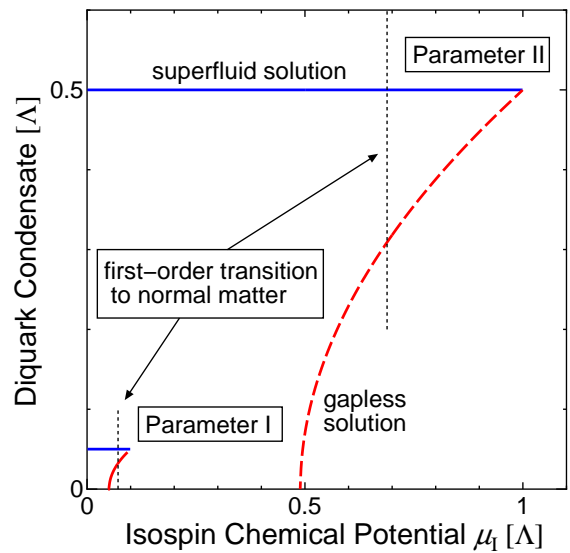}
 \caption{Superfluid and gapless solutions at $\muq=0.5\Lambda$.  The
 first-order phase transition from superfluid to normal quark matter
 is indicated by the dotted line.}
 \label{fig:gapless}
\end{figure}

  Once $\muI$ becomes larger than $\Delta$, $\epsilon_{\text{p}}^-$
decreases down to zero at $p\sim\muq$, that is, the energy dispersion
relation become gapless.  This gapless superfluid solution is similar
to the gapless 2SC (g2SC) phase known from QCD.\ \ In contrast to the
g2SC case, we do not impose neutrality on the present system.  In
principle, we can consider electric neutrality by turning the
electromagnetic interaction on and putting electrons into the system.
We will not do so, however, partly because the resultant stability of
the gapless state with respect to homogeneous change in the gap
amplitude as seen in the g2SC case would not be of interest here, and
partly because we intend to keep an interesting resemblance of the
present system with electric charge turned off and $\muI$ introduced
to atomic Fermi gases with population imbalance created in low
temperature experiments.  Note that even for charged systems, the
gapless superfluid state is generally unstable with respect to
inhomogeneous fluctuations in the order parameter~\cite{Iida:2006df}.

  The typical behavior of the diquark condensate as a function of
$\muI$ is plotted in Fig.~\ref{fig:gapless} for the weak and
intermediate coupling cases.  Both are the results at
$\muq=0.5\Lambda$ and thus the superfluid solution stays at
$\Delta=0.05\Lambda$ for Parameter~I and at $\Delta=0.5\Lambda$ for
Parameter~II in the figure.  There appear two branches in the
solutions to Eq.~(\ref{eq:gapeq}) for a finite range of $\muI$,
namely, the gapless solution and the superfluid solution.  When these
two solutions meet at a certain $\muI$, the gap equations cease to
have any solution.  It happens at $\muI=0.099\Lambda$ for Parameter~I
and $\muI=0.99\Lambda$ for Parameter~II in Fig.~\ref{fig:gapless}.  As
we shall see from the energy landscape, the thermodynamic potential
$\Omega$ has an inflection point when one solution corresponding to a
local minimum of the potential meets the other corresponding to a
local maximum.  This means that there must exist another state which
takes over the ground state before $\muI$ reaches the meeting point.
In fact, a first-order phase transition to normal matter is found at
$\muI=0.0355\Lambda$ or, equivalently,
\begin{equation}
 \muI = 0.711\Delta_0 \quad \text{(for Parameter~I)}\;,
\end{equation}
where $\Delta_0=0.05\Lambda$ at weak coupling, and
$\muI=0.344\Lambda$ or, equivalently,
\begin{equation}
 \muI = 0.688\Delta_0 \quad \text{(for Parameter~II)}\;,
\end{equation}
where $\Delta_0=0.5\Lambda$ at intermediate coupling.  As discussed in
Ref.~\cite{Alford:2000ze}, this $\muI$ of the first-order transition
approximately gives the lower bound $\delta\mu_2$ of the LOFF favored
region if the upper bound $\delta\mu_1$ of the LOFF solution, as will
be discussed later, is greater than $\delta\mu_2$.  The point is that
the energy gain in the LOFF phase relative to normal quark matter is
tiny as compared with the scale of the change in the energy difference
between the superfluid and normal phases with increasing $\muI$.  In
order for the LOFF phase to be most favorable, therefore, it is
necessary that the energy gain in the superfluid phase be very close
to zero, which is only possible when $\muI$ is very close to the
first-order phase transition point.  We note that our weak coupling
value of $\delta\mu_2=0.711\Delta_0$ precisely agrees with the known
result in Ref.~\cite{Alford:2000ze}.

%%%   LOFF Phase   %%%

\subsection{LOFF Phase}

  Let us now move on to the central part of this paper that addresses
the LOFF favored region in $\muq$-$\muI$ space.  Here we shall discuss
the weak and intermediate coupling cases separately.

  In the presence of nonzero $\q$, the free energy has an unphysical
term proportional to $-\Lambda^2 q^2$ which must be subtracted.  It is
a nontrivial problem how to renormalize this spurious contribution
properly in a field-theoretical procedure.  [See discussions in
Refs~\cite{Alford:2005qw,Fukushima:2005cm} about this problem in
evaluating the Meissner mass.]  Our preference is to follow a
practical prescription here.  After we solve the gap equations with a
given $\q$ to get $\sigma=\sigma(\Delta,q)$, which is not affected by
the term $\sim-\Lambda^2 q^2$ in question, we can define the
subtracted free energy as
\begin{equation}
 \Omega_{\text{s}}(\Delta,q)
 = \Omega[\sigma(\Delta,q),\Delta(q),q]
  - \Omega[\sigma(\Delta,q),0,q] \;,
\label{eq:s_energy}
\end{equation}
in order to determine the gap $\Delta$ and the optimal value of $q$
from its global minimum.  This prescription is the simplest choice
consistent with the fact that $\Omega_{\text{s}}$ should be flat in
the $q$-direction in the absence of finite $\Delta$.  The second
term in the right-hand side could be $-\Omega[\sigma(0,q),0,q]$, but
it would make only a negligible difference in numerical outputs.

%--   Weak coupling case   ---%

\subsubsection{Weak coupling case}

\begin{figure}
 \includegraphics[width=7.5cm]{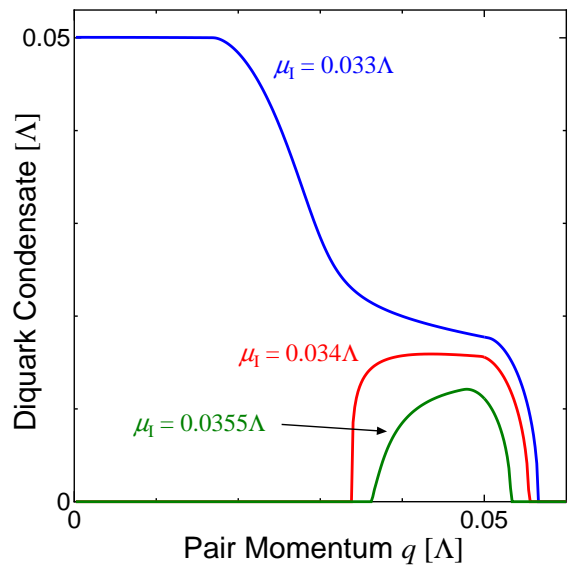}
 \caption{Diquark condensate at weak coupling as a function of the
 pair momentum $q$, which is calculated for $\muI=0.033\Lambda$ (top), 
 $0.034\Lambda$ (middle), and $0.0355\Lambda$ (bottom).}
 \label{fig:loff_w}
\end{figure}

\begin{figure}
 \includegraphics[width=7.5cm]{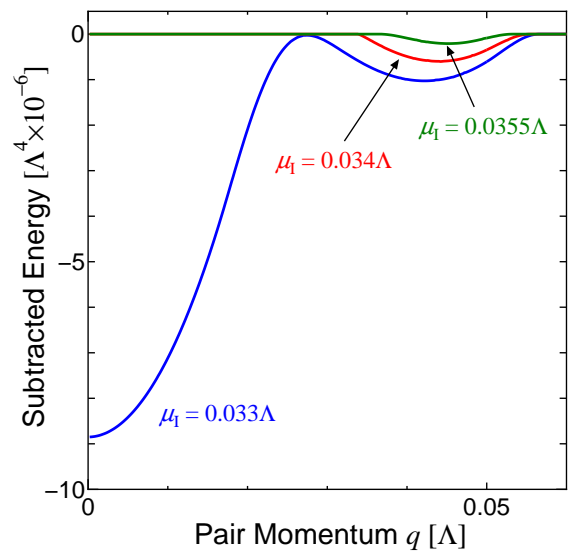}
 \caption{Energy difference (\ref{eq:s_energy}) at weak coupling as a 
function of the pair momentum $q$.   The three curves correspond to 
those displayed in Fig.~\ref{fig:loff_w}.}
 \label{fig:loff_energy_w}
\end{figure}

  We can solve the gap equation
$\partial\Omega_{\text{s}}/\partial\Delta=0$ to obtain
$\Delta=\Delta(q)$ as a function of $q$.  Figure~\ref{fig:loff_w}
shows the numerical results for $\Delta(q)$ with Parameter~I (weak
coupling) for several values of $\muI$ in the vicinity of the critical
value $\muI=\delta\mu_2$.  We can observe from the curve labeled with
$\muI=0.033\Lambda$ that $\Delta(q)$ is smoothly connected up to the
value at $q=0$ which corresponds to the diquark condensate in the
superfluid phase.  It is clear from $\Omega_s$ plotted in
Fig.~\ref{fig:loff_energy_w} that, when $\muI=0.033\Lambda$, the
superfluid phase at $q=0$ has a much smaller energy than the
metastable LOFF solution at $q\neq0$.  Interestingly the LOFF solution
becomes disconnected from the superfluid solution as $\muI$ grows up.
Nevertheless, the superfluid phase continues to exist separately and
is more favorable until the LOFF phase turns to the ground state at
$\muI>\delta\mu_2$.

  Then, there occurs a crucial question:  How large is the upper bound
$\delta\mu_1$ of $\muI$ at which the LOFF solution disappears?  The
$q$-range of nonzero $\Delta(q)$ shrinks as $\muI$ gets larger as
indicated in Fig.~\ref{fig:loff_w}.  We find that it is located at
$\muI=0.0377\Lambda$ or, equivalently,
\begin{equation}
 \delta\mu_1 = 0.755 \Delta_0 \;,
\end{equation}
where $\Delta_0=0.05\Lambda$, and that the optimal pair momentum at
$\muI=\delta\mu_1$ reads $q=0.045\Lambda$ or, equivalently,
\begin{equation}
 q \simeq 1.2\delta\mu_1 \;.
\end{equation}
Our $\delta\mu_1$ is slightly bigger than the value reported in
Ref.~\cite{Alford:2000ze}.  We expect that the discrepancy could be
reduced if we choose smaller $\Delta_0$ and smaller $\muq$.

  It would be intriguing to turn our focus toward a even higher
density region where the saturation effect is relevant.  Then, the
above result for the LOFF favored region,
$\delta\mu_1=0.755\Delta_0 >\muI> \delta\mu_2=0.711\Delta_0$, may well
be altered, and indeed the LOFF window enlarges a bit.

\begin{figure}
 \includegraphics[width=7.5cm]{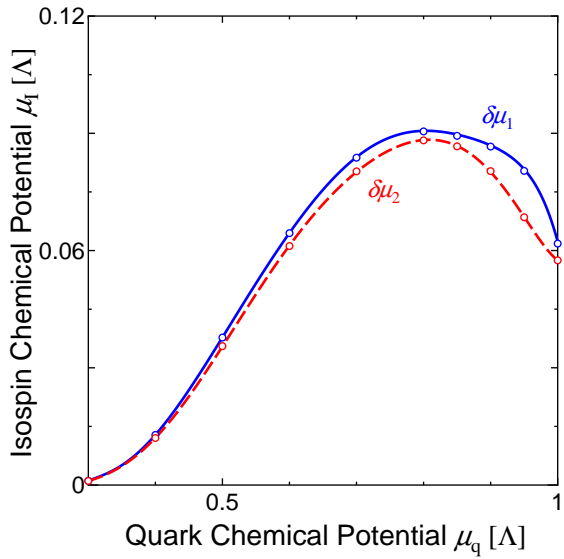}
 \caption{The LOFF favored region bounded by $\delta\mu_1$ and
 $\delta\mu_2$, calculated at weak coupling as a function of $\muq$.
 The open circles represent the calculated values, whereas the curves
 are interpolations between the neighboring circles by splines.}
 \label{fig:loff_phase_w}
\end{figure}

  Figure~\ref{fig:loff_phase_w} demonstrates that the LOFF window
varies with changing quark chemical potential $\muq$.  The shape of
the curves reflects that of $\Delta_0(\muq)$ given in
Fig.~\ref{fig:cond}.  The LOFF window becomes slightly wider in the
region where the saturation effect is sufficient to make the curves
have a negative slope.  For instance, at $\muq=0.95\Lambda$ the LOFF
favored region is bounded by
\begin{equation}
 \begin{split}
 \delta\mu_1 &= 0.080\Lambda = 0.73\Delta_0(\muq\!=\!0.95\Lambda) \;, \\
 \delta\mu_2 &= 0.069\Lambda = 0.62\Delta_0(\muq\!=\!0.95\Lambda) \;,
 \end{split}
\end{equation}
where $\Delta_0=0.110\Lambda$.  The interval between them is not
substantial in the weak coupling case, while the behavior drastically
changes at intermediate coupling as we will see shortly.

%---   Intermediate coupling case   ---%

\subsubsection{Intermediate coupling case}

\begin{figure}
 \includegraphics[width=7.5cm]{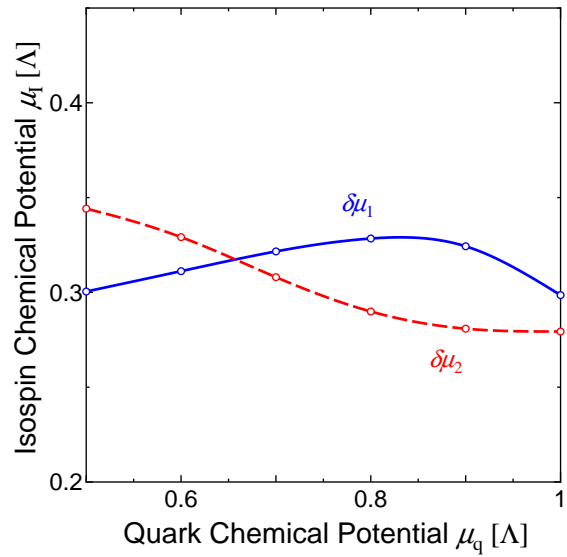}
 \caption{Same as Fig.~\ref{fig:loff_phase_w} except that calculations
 are performed at intermediate coupling.}
 \label{fig:loff_phase_s}
\end{figure}

  In the intermediate coupling case with Parameter~II, the gross
behavior of the diquark condensate and the subtracted energy as a
function of $q$ is just similar to what we have shown in
Figs.~\ref{fig:loff_w} and \ref{fig:loff_energy_w}.  We thus find that
such behavior has only a weak dependence on the parameter choice.  For
the LOFF favored region, however, the parameter dependence is
important as can be seen from comparison between
Figs.~\ref{fig:loff_phase_w} and \ref{fig:loff_phase_s}.

  The first-order transition from the superfluid to the normal phase
occurs above the upper bound of the LOFF solution for
$\muq\lesssim 0.7\Lambda$.  This means that the LOFF window shuts
there and the superfluid phase remains the ground state even for
$\muI>\delta\mu_1$ until normal quark matter appears.  This behavior,
which is a contrast to the case with Parameter~I, is caused by
stronger coupling rather than by heavier $m_0$.  In fact, the chiral
condensate is much smaller than the diquark condensate at
$\muq\sim 0.7\Lambda$ as shown in Fig.~\ref{fig:cond}.

  The LOFF window opens at $\muq\gtrsim 0.7\Lambda$ where we find
the same tendency as the weak coupling case that the LOFF window
widens by the saturation effect.  At $\muq=0.8\Lambda$, for instance,
the bounds of the LOFF window are given by
\begin{equation}
 \begin{split}
  \delta\mu_1 &= 0.33\Lambda = 0.64\Delta_0(\muq\!=\!0.8\Lambda) \;,\\
  \delta\mu_2 &= 0.29\Lambda = 0.57\Delta_0(\muq\!=\!0.8\Lambda) \;,
 \end{split}
\end{equation}
where $\Delta_0=0.51\Lambda$ at $\muq=0.8\Lambda$.

  We note that our plane-wave LOFF ansatz~(\ref{eq:loffansatz}) is
only the simplest one.  If we consider more general three-dimensional
crystalline structure, the LOFF favored region has to be even wider.
We would stress that our results shown in Fig.~\ref{fig:loff_phase_s}
strongly support the possibility of probing the generic crystalline
color superconducting phase, along with the isotropic superfluid
phase, with numerical approach on lattice.  In fact, the values of
$\muI$ at which the LOFF phase occurs are less than a half of $\muq$,
while being larger than the inverse of a typical lattice size.
Possible simulation to detect the LOFF phase would not require weaker
coupling, finer lattice, or larger lattice.

%%%   Energy Landscape   %%%

\subsection{Energy Landscape}

\begin{figure}
 \includegraphics[width=10cm]{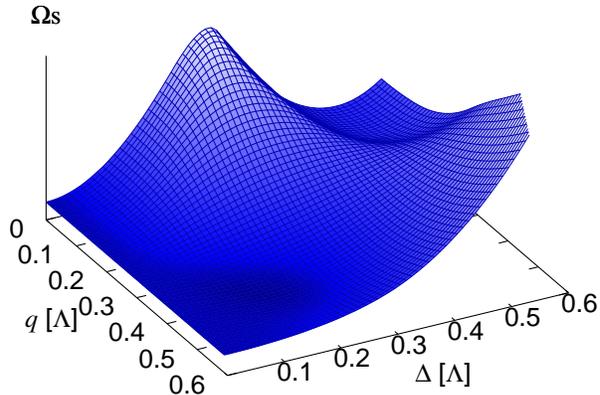}
 \caption{Subtracted free energy as a function of $\Delta$ and $q$
 with $M=M(\Delta,q)$ which is the solution to the gap equation.  A
 darker colored region has a smaller energy.  The LOFF phase is the
 global minimum of the energy.}
 \label{fig:energy}
\end{figure}

  We will take a further look at the LOFF phase found at intermediate
coupling.  Let us pick up a point of $\muq=0.8\Lambda$ and
$\muI=0.3\Lambda$ inside the LOFF favored region illustrated in
Fig.~\ref{fig:loff_phase_s}.  In order to elucidate how the unstable
gapless phase is connected to the LOFF phase, we picture the energy
landscape, namely, we plot $\Omega_{\text{s}}$ as a function of
$\Delta$ and $q$ as shown in Fig.~\ref{fig:energy}.

  It is obvious from Fig.~\ref{fig:energy} that the LOFF phase sits
certainly at the global minimum of the potential energy landscape.
The local minimum at $\Delta=q=0$ is normal quark matter, another
local minimum at $\Delta\neq0$ and $q=0$ is the metastable superfluid
phase, and the in-between local maximum along the $\Delta$-axis
corresponds to the unstable gapless phase.  The gapless phase is
unstable with respect to fluctuations in any direction in the space of
$\q$ and $\Delta$.  Specifically, the instability along the $\Delta$
direction is the Sarma instability originally noticed in
Ref.~\cite{Sarma}.  In the g2SC or gCFL phase, the remedy against the
Sarma instability comes from the electric and color neutrality
conditions.  Even with neutrality imposed, however, the instability in
the $q$ direction is still a problem, which is nothing but the
chromomagnetic instability in the g2SC or gCFL phase.  For a neutral
superfluid of interest here, it is the phase instability.

\begin{figure}
 \includegraphics[width=8.7cm]{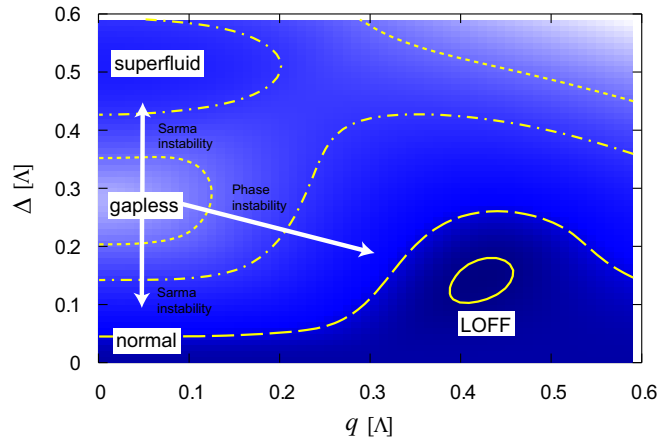}
 \caption{Contour density plot of $\Omega_{\text{s}}$, which indicate
 two minima corresponding to the metastable superfluid phase and the
 ground state of the LOFF phase and one maximum corresponding to the
 unstable gapless phase.}
 \label{fig:contour}
\end{figure}

  Figure~\ref{fig:contour} is the contour density plot of the
numerical data in Fig.~\ref{fig:energy}, which allows us to describe
graphically what we mentioned above.  The gapless phase has
instabilities leading to normal quark matter, the isotropic superfluid
phase, and the LOFF phase, among which the LOFF phase has the lowest
energy.  This contour plot presumably captures the essence of the
instability problem and its implications for QCD quark matter.  We
would anticipate that the energy landscape around the g2SC phase is
more or less similar to Fig.~\ref{fig:contour} except that the gapless
phase is stabilized against fluctuations in $\Delta$, that is, the
gapless state is a saddle point on the energy landscape.  It is hard,
however, to imagine what would happen in the case of the gCFL phase
with three flavors,  because there are three predominant diquark
condensates.  In large dimensional space spanned by more variational
parameters, it is not straightforward to see whether the instability
in the gapless phase is directly connected to the LOFF phase.  There
could be several distinct LOFF solutions as conjectured in
Ref.~\cite{Fukushima:2006su}.  It would be a challenging problem to
draw the energy landscape in the three-flavor case and to confirm that
the instability is responsible for the LOFF phase, as we have done
successfully in two-color and two-flavor QCD in this work.

%%%%%%%%%%   FINITE TEMPERATURES   %%%%%%%%%%

%\section{FINITE TEMPERATURES}

%%%%%%%%%%   CONCLUSIONS   %%%%%%%%%%

\section{CONCLUSIONS}
\label{sec:conclusion}

  We investigated the LOFF favored window of $\muI$ for two-color and 
two-flavor QCD using a mean-field model.  First, we satisfactorily
reproduced the known results at weak coupling.  We then performed our
calculation for a set of the model parameters in which the coupling
strength is intermediate and thus relevant to currently available
lattice simulations.  We found out a nonzero interval of $\muI$ where
the single plane-wave LOFF phase is energetically more favorable than
the isotropic superfluid phase and normal quark matter.  For
intermediate coupling, we took a close look at the energy landscape to
see the relation of the Sarma and phase instabilities associated with
the gapless solution of the gap equations with the LOFF phase.

  In this work we do not take account of any possibility of the mixed
phase or the phase separation for a nonzero range of $\muq$ as
discussed in Ref.~\cite{Iida:2006df} because we turn electric charge
off in the present system.  The chargeless limit would be advantageous
to lattice two-color QCD simulation.  In the absence of the fermion
sign problem, furthermore, we believe that lattice approach to the
LOFF phase is feasible~\cite{future}.

\begin{acknowledgments}
We thank Atsushi Nakamura and Chiho Nonaka for discussions.
K.~F.\ thanks Masakiyo Kitawaza for comments.  He also thanks Sinya
Aoki, Yasumichi Aoki, and Shinji Ejiri for
conversations.  This research was supported in part by RIKEN BNL
Research Center and the U.S.\ Department of Energy under cooperative
research agreement \#DE-AC02-98CH10886.
\end{acknowledgments}

%%%%%%%%%%   REFERENCES   %%%%%%%%%%

\end{document}